\documentclass[%
 aip,
 amsmath,amssymb,
reprint,%
pop,
]{revtex4-1}

\usepackage{graphicx}
\usepackage{dcolumn}
\usepackage{bm}

\usepackage[utf8]{inputenc}
\usepackage[T1]{fontenc}
\usepackage{mathptmx}
\usepackage{etoolbox}
\usepackage{siunitx}
\usepackage{xcolor}
\usepackage{placeins}
\usepackage{textgreek}
\usepackage{amsmath}
\usepackage{hyperref}
\usepackage{cleveref}

\makeatletter
\def\@email#1#2{%
 \endgroup
 \patchcmd{\titleblock@produce}
  {\frontmatter@RRAPformat}
  {\frontmatter@RRAPformat{\produce@RRAP{*#1\href{mailto:#2}{#2}}}\frontmatter@RRAPformat}
  {}{}
}%
\makeatother
\begin{document}

\preprint{AIP/123-QED}

\title{Collisionless Larmor Coupling and Blob Formation in a Laser-Plasma Expanding into a Magnetized Ambient Plasma}

\author{L. Rovige}
\affiliation{University of California--Los Angeles Los Angeles, CA 90095, USA}
\affiliation{Laboratoire d’Optique Appliquée (LOA), CNRS, École polytechnique, ENSTA, Institut Polytechnique de Paris, Palaiseau, France}

\author{R. S. Dorst}
\affiliation{University of California--Los Angeles Los Angeles, CA 90095, USA}
\affiliation{Lawrence Livermore National Laboratory, 7000 East Avenue, Livermore, California 94550, USA}

\author{A. Le}
\affiliation{Los Alamos National Laboratory, Los Alamos, New Mexico 87544, USA}

\author{C. G. Constantin}
\affiliation{University of California--Los Angeles Los Angeles, CA 90095, USA}

\author{H. Zhang}
\affiliation{University of California--Los Angeles Los Angeles, CA 90095, USA}

\author{D. J. Larson}
\affiliation{Lawrence Livermore National Laboratory, 7000 East Avenue, Livermore, California 94550, USA}

\author{S. Vincena}
\affiliation{University of California--Los Angeles Los Angeles, CA 90095, USA}

\author{S.K.P Tripathi}
\affiliation{University of California--Los Angeles Los Angeles, CA 90095, USA}

\author{M. Cowee}
\affiliation{Los Alamos National Laboratory, Los Alamos, New Mexico 87544, USA}

\author{D. B. Schaeffer}
\affiliation{University of California--Los Angeles Los Angeles, CA 90095, USA}

\author{C. Niemann}
\affiliation{University of California--Los Angeles Los Angeles, CA 90095, USA}
\date{\today}

\begin{abstract}

Collisionless Larmor coupling is a fundamental process in space and astrophysical plasmas that enables momentum transfer between an expanding plasma and a magnetized ambient medium. In this paper, we report on the laboratory experimental study of Larmor coupling leading to the formation of a plasma blob associated with a laser-driven, super-Alfvénic plasma flow on the Large Plasma Device at the University of California, Los Angeles. The high-repetition rate enables systematic spatial and temporal scans of the plasma evolution using Doppler spectroscopy, as well as measurements of the magnetic field, electrostatic field, and self-emission of both debris and ambient ions using filtered imaging.  We observe the self-focusing of the laser-produced plasma and the formation of a secondary diamagnetic cavity associated with a blob composed of background ions. Doppler spectroscopy reveals the transverse velocity distribution of the background ions, providing direct evidence of ion energization via Larmor coupling. The systematic spatial and temporal scans enabled by the high-repetition rate experiment allow for a detailed characterization of the ion dynamics. These experimental observations are supported by numerical simulations that provide more insight into the kinetic-scale physics associated with blob formation as well as the role of the ambient plasma density. 
\end{abstract}

\maketitle

\section{Introduction}

The interaction of an expanding plasma cloud with a magnetized background is a fundamental process in space and astrophysical plasmas that can occur across vastly different scales, from the expansion of supernova remnants into the interstellar medium \cite{spicer90} to the formation of cometary plasma tails \cite{Mendis1977} and high-altitude nuclear explosions \cite{Palmer06}. In these rarefied environment, the collisionless interaction between the expanding plasma and the magnetized ambient can lead to the formation of plasma blobs—localized, high-density structures that exhibit coherent cross-field motion\cite{Chaturvedi1987,Huba1992,Park2003,Katz2008,Theiler09,Mackey2024}. Such blobs have been observed in space associated with ionospheric disturbances \cite{Colgate1965,Zinn1966,Dyal2006}, solar flares \cite{Chen2022}, as well as cometary transits \cite{Kim_2023} and in the laboratory in fusion devices and other magnetized experiments\cite{Nielsen1996,Endler1999,Boedo2003,Terry2003,Carter2006,Fasoli2006, Furno2008} and laser-driven plasmas. \cite{Bondarenko2017,Dorst2025}

Plasma blobs can generally occur in two types of conditions that can be categorized either as laminar - where the momentum transfer occurs through large scale polarization and induction electric fields, or turbulent - where the blob formation is associated with the development of instabilities at the interface between the expanding plasma and the magnetized ambient medium \cite{Lampe1975,spicer90}. In the laminar regime, the formation of plasma blobs is generally explained by the development of strong polarization electric fields, which in turn drive $\mathbf{E}\times\mathbf{B}$ drift currents. This process can result in the formation of blobs that propagate across field lines. This description works well for large systems evolving on fluid scales, but does not account for finite Larmor radius effects that arise on kinetic spatial and temporal scales, where it has been demonstrated that Larmor coupling can play a significant role in the momentum transfer between the debris and ambient plasma \cite{Golubev1978,Bashurin1983}.

This work aims to reproduce space-like conditions characterized by a super-Alfvénic flow (\( M_A = v_d / v_A > 1 \), where \( v_d \) is the debris velocity and \( v_A \) is the Alfvén speed in the ambient plasma) and a magnetically dominated regime (\( \beta = 8\pi p_e / B^2 \ll 1 \), with \( p_e \) the electron thermal pressure). In this regime, turbulence has been shown to be relatively inefficient at transferring momentum~\cite{Papadopoulos1971,McBride1972}, and instead, momentum transfer is expected to be mediated by large-scale electric fields. In this low-$\beta$, cross-field flow regime, the laminar electric field can be well approximated by\cite{Bondarenko2017_2}:

\begin{equation}
E_{\text{lam}} \simeq - \frac{\mathbf{B} \times (\nabla \times \mathbf{B})}{4\pi e \sum_i Z_i n_i} - \frac{\sum_i Z_i n_i \mathbf{v}_i \times \mathbf{B}}{c \sum_i Z_i n_i}
\end{equation}
Where $v_i$ is the ion velocity, $n_i$ is the ion density, $e$ is the electron charge, and $Z_i$
is ion charge. This left-hand term arises from magnetic pressure gradients in the plasma, and points towards decreasing magnetic field gradients. The right-hand term represent the electric field induced by ions current transverse to the magnetic field and is denominated Larmor electric field. 

Observations from space missions such as the Combined Release and Radiation Effects Satellite (CRRES) \cite{Huba1992,Huba1993} and the Active Magnetospheric Particle Tracer Explorer (AMPTE) \cite{Bernhardt1989,Krimigis1982} have documented the emergence of plasma blobs following the release of artificial ion clouds in the upper atmosphere, demonstrating their ability to persist and transport across field lines. The formation and transport of these blobs were attributed to Larmor coupling \cite{Golubev1978,Bashurin1983}, a mechanism in which the background ions are initially accelerated transversely by the Larmor electric field. These ions then undergo gyromotion in the magnetic field, effectively transferring their gained momentum into forward motion and enabling them to be picked up by the expanding plasma debris. These predictions have been corroborated by laboratory experiments from our group, which observed evidence of Larmor coupling between a laser-produced plasma and a magnetized ambient plasma\cite{Bondarenko2017}. A more recent experiment provided a deeper insight into the laser-plasma self-focusing and blob formation, with a particular focus of the dynamics of the debris plasma using laser-induced fluorescence (LIF) on the laser-produced carbon ions \cite{Dorst2025}.

In this work, we present additional experimental data using a wide range of diagnostics providing a more complete picture of the collisionless coupling between the LPP and the ambient plasma leading to the formation and acceleration of a plasma blob. We notably study Larmor coupling using Doppler spectroscopy in order to access the transverse velocity distribution of the background ions, in a more systematic way than before \cite{Bondarenko2017,Bondarenko2017_2} through spatial and temporal scans enabled by the high-repetition rate of the experiment. Additionally, a background plasma density twice as high as in previous experiments was achieved, providing new insights into the role of density in the blob formation process. The experimental setup and diagnostics are presented in Sec. \ref{sec:setup} and the obtained results in Sec. \ref{sec:res}. Particle-in-Cell (PIC) simulations and associated physical discussions are presented in Sec. \ref{sec:PIC}.

\section{Experimental Setup}

\, \newline
\begin{figure}
\includegraphics[width=0.98\linewidth]{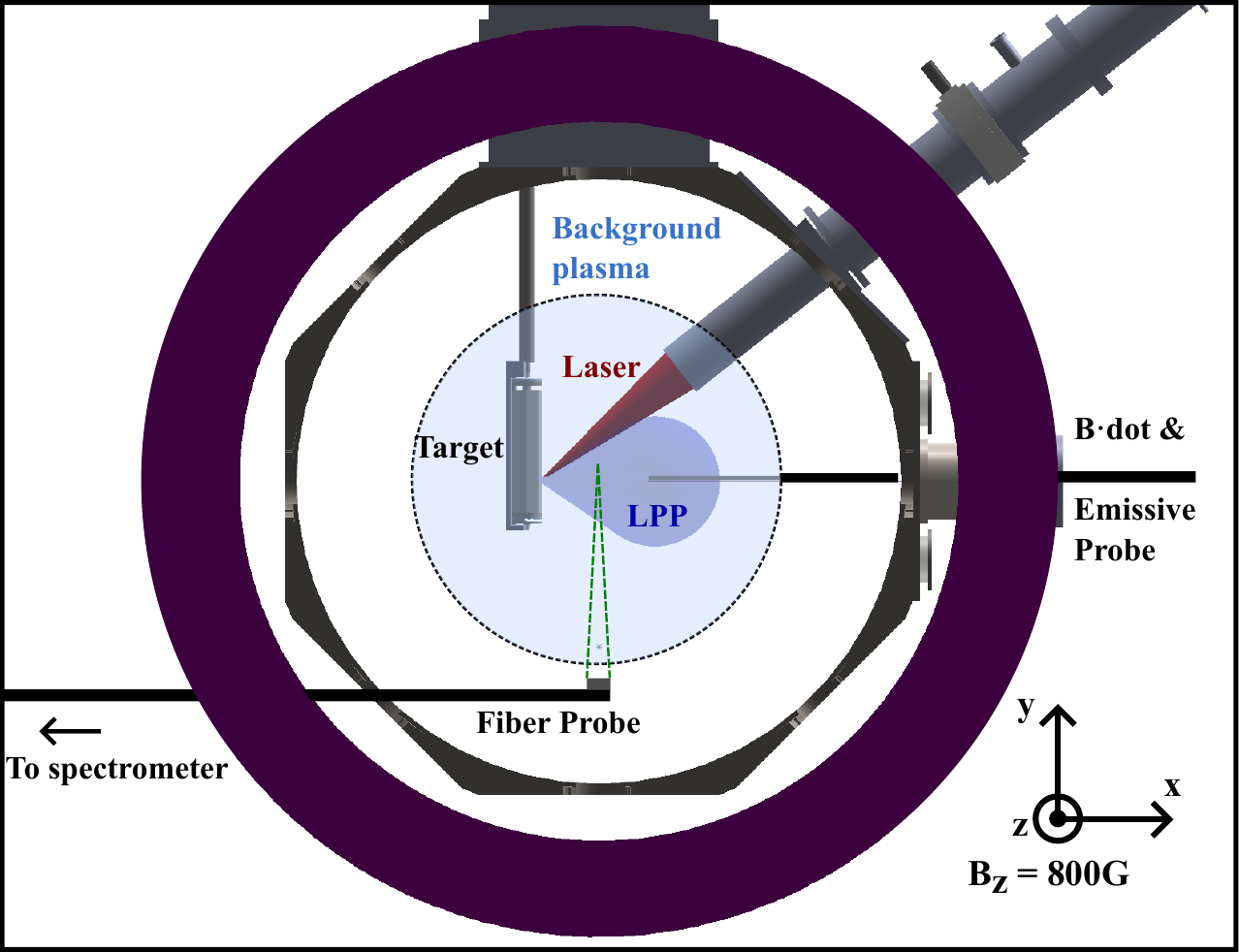}
\caption{\label{fig:fig1} Schematic of the experimental setup on the LAPD.}
\end{figure}

\label{sec:setup}
The experiment was conducted on the Large Plasma Device (LAPD), operated by the Basic Plasma Science Facility (BaPSF) at UCLA, and utilized a high-power laser system from the Phoenix Laser Facility to investigate the interaction between a laser-driven, super-Alfvénic plasma and the magnetized ambient plasma.

The LAPD \cite{Gekelman2016} is a linear, pulsed discharge device comprising a 20-meter-long, 1-meter-diameter vacuum vessel capable of generating a long-lived ($\sim 20$ ms), highly-repeatable plasma column at repetition rates of up to 1 Hz. Circular magnetic coils placed regularly along the length of the machine generate an axial (along $z$) magnetic field adjustable between 200~G and 2500~G. The plasma density can be varied within the range of $10^{12} - 5\times 10^{13} \,\text{cm}^{-3}$, with typical electron temperatures of $T_e \sim 2 - 10$ eV and ion temperatures around $T_i \sim 1$ eV.
The laser, operating at a repetition rate of 1\,Hz, delivers 10\,J of energy in a 20\,ns pulse at $\lambda = 1053$\,nm. It is used to generate a super-Alfvénic plasma flow by focusing the beam onto a cylindrical graphite target, achieving an intensity of $I \sim 10^{12} \,\text{W/cm}^2$. To ensure a fresh surface for each laser shot, the motorized target is rotated in a helical pattern, enabling up to 12,000 consecutive shots.  The laser-produced plasma is mainly constituted from $C^{4+}$ ions \cite{Schaeffer2016} and propagates along the x-axis at a speed of $v_{lpp} = $ 270\,km/s, transverse to the background magnetic field. The ion-ion collision mean-free path between the carbon and helium ions is $\lambda_{ii} = 22$~m so that their interaction is largely collisionless on the scale of our experiment.

In this experiment, the LAPD is filled with helium gas, producing a background plasma composed of He$^+$ ions. Helium is chosen as it enables imaging and spectroscopy, as well as reaching a higher density. The background plasma density is measured with a combination of a Langmuir probe and 288\,GHz interferometer and was $n_e = n_{He^+} = 4\times10^{13}$\,cm$^{-3}$, yielding an inertial length for the $He^+$ ions $d_i = c/\omega_{pi} = 7$\,cm where $\omega_{pi} = (n_i Z^2 e^2/(\varepsilon_0 m_i))^{1/2}$ is the ion plasma frequency. The background magnetic field is set to be $B_0 = 800$\,G with  $\mathbf{B_0} = B_0 \mathbf{u_z}$, yielding a super-Alfvénic flow with a Mach number $M_A = 2.1$. 

\begin{figure*}[ht!]
\includegraphics[width=0.85\linewidth]{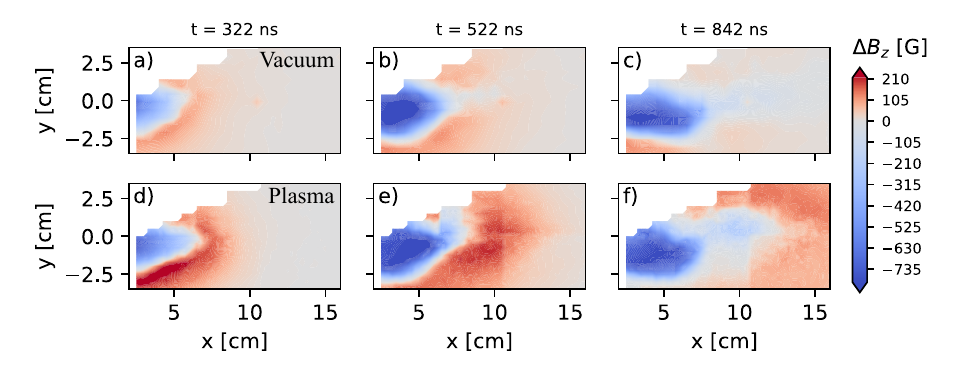}
\caption{\label{fig:fig2} Variation of the z component of the magnetic field $\Delta B_z$ relative to the initial background field, for: a)-c) An expansion of the LPP in vacuum d)-f) An expansion of the LPP in the background plasma. The probe could not access the top left corner of the plane, as this region intersected the laser beam path.  }
\end{figure*}

The time-resolved magnetic field and plasma potential are measured in the XY-plane using a magnetic flux (B-dot)\cite{Everson2009} and emissive probe\cite{Martin2015}. The quantities are measured locally with three repetitions per position, and the probes moved between shots to map out the fields in 2D. An intensified (ICCD) PI-MAX4 camera with up to 4~ns time resolution images the interaction and captures snapshots at different delays on different shots. Narrow band-pass filters were placed in front of the camera, using either a 468.6\,nm filter to image emission from an excited level of He$^{+}$ or a 227\,nm filter to image emission from the C$^{4+}$ ions. A spectroscopic fiber-probe is inserted from the side, below the target, collecting emitted light from the plasma line integrated in the z-direction. A 75mm focal length lens projects the image of the plasma onto a line array bundle of 20 -- 200\,$\mu$m optical fibers, yielding a field of view of $\sim$1.5\,cm in the z-direction and 0.1\,cm in the x direction. The fiber-probe is motorized so that it can scan and collect light at different positions along the x-axis. The collected light is sent to a grating spectrometer centered on the 468.6\,nm excited He$^+$ line and imaged on a fast-gate, intensified (ICCD) PI-MAX2 camera, resulting in a spectral resolution of 0.02\,nm.

\section{Experimental results}
\label{sec:res}

\begin{figure*}[ht!]
\includegraphics[width=1\linewidth]{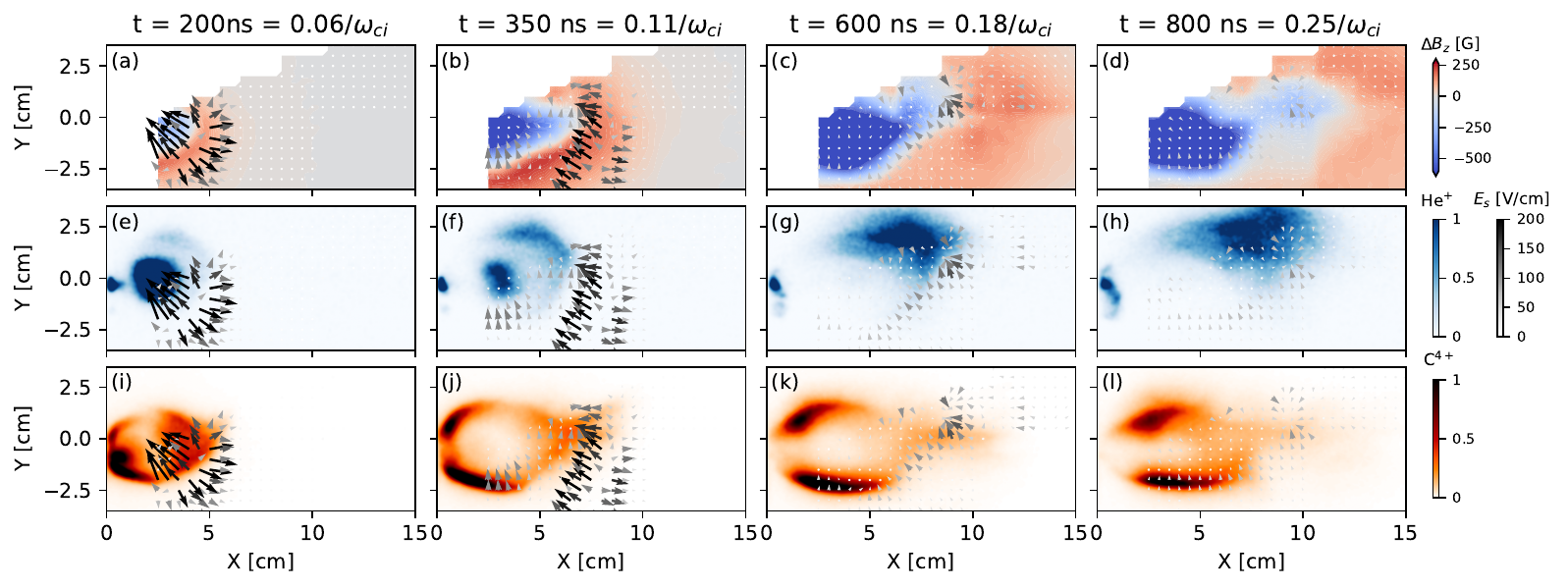}
\caption{\label{fig:fig3} (a)-(d) Variation of the magnetic field measured with a B-dot probe, (e)-(h) Filtered imaging of the plasma using a bandpass filter at 468.6\,nm with a 0.5\,nm bandwidth, corresponding to an excited He$^{+}$ emission line, with an exposure time of 50\,ns (i)-(l) Filtered imaging at 227\,nm with a 10\,nm bandwidth, corresponding to a C$^{4+}$ emission line, with an exposure time of 4\,ns. The direction of the electrostatic field, derived from plasma potential measurements using the emissive probe, is depicted by arrows whose length and color indicate the field's magnitude. }
\end{figure*}

As the fast-moving laser plasma expands into the background magnetic field, it sweeps the field away to create a diamagnetic cavity, while compressing it to the front of its expansion \cite{Niemann2013, Schaeffer2018}. Figure \ref{fig:fig2} shows the evolution of this diamagnetic cavity both in vacuum and plasma background. The cavity experiences collimation as it is pinched from the sides by the magnetic pressure associated with the large gradients of the magnetic field on the edges of the laser-plasma. This phenomenon has been characterized in a previous experiment measuring the evolution of the transverse velocity of the $C^{4+}$ ions using laser-induced florescence (LIF) \cite{Dorst2025}. In the background plasma case, a secondary diamagnetic cavity forms and expands from the upper part of the tip of the laser-produced plasma around $t \sim 500$\,ns. Later in time, as the main diamagnetic cavity's expansion stagnates at $x = 7.5$\, cm, this secondary structure detaches and continues propagating forward.

\begin{figure}[ht]
\includegraphics[width=1\linewidth]{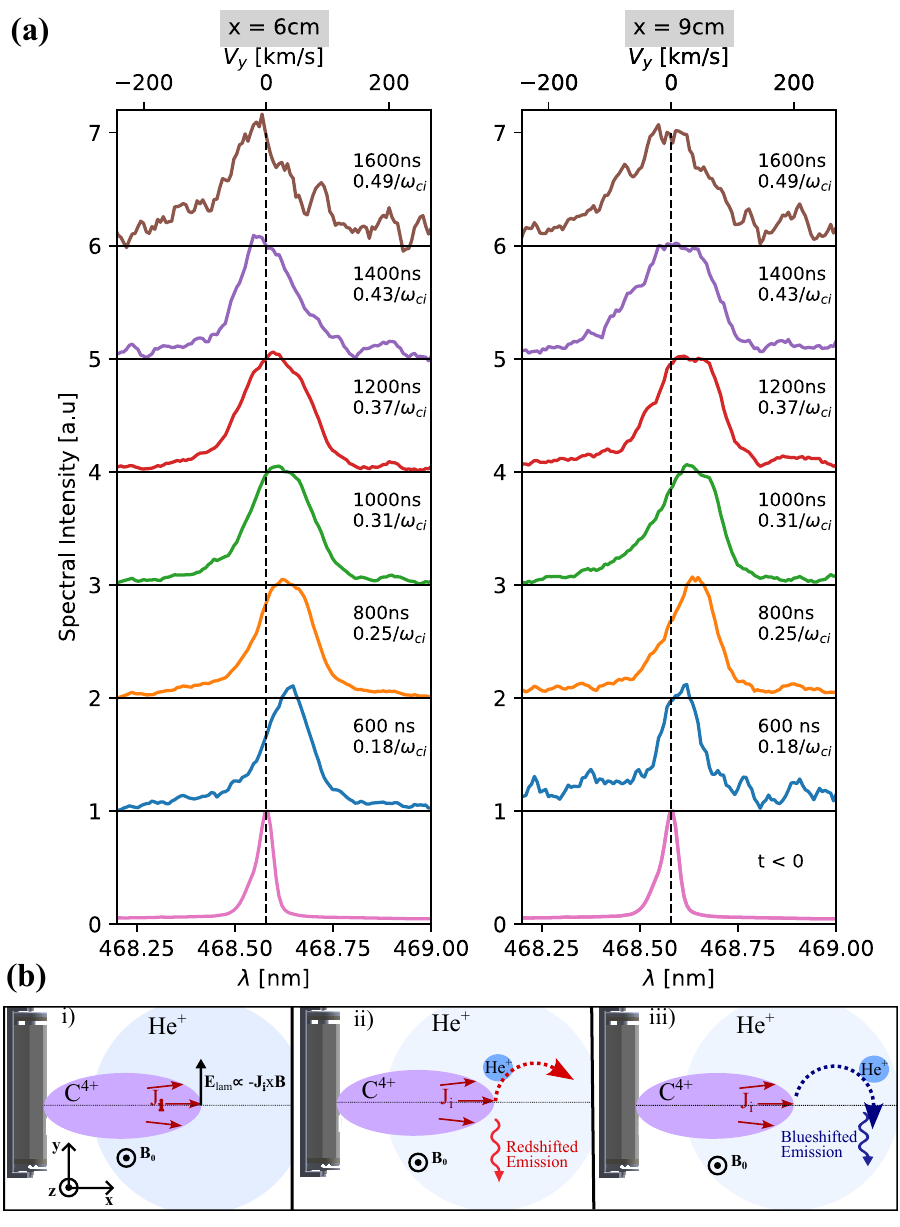}
\caption{\label{fig:fig4} (a) Doppler-shifted spectra and corresponding transverse velocity $V_y$ at different times at x = 6\,cm (left) and x = 9\,cm (right). The $t<0$ case corresponds to the He$^+$ line measured in the resting background plasma. It provides the impulse response of the spectrometer, with a full-width half max of $\delta\lambda=$ 0.04\,nm, corresponding to a velocity resolution $\delta V_y =$ 25\,km/s. The spectra were acquired with an exposure time of 200\,ns and averaged on 200 shots.  (b) Principle of Larmor coupling and its Doppler shift signature.}
\end{figure}

Figure \ref{fig:fig3} shows magnetic field data alongside filtered images of self emission of He$^+$ and C$^{4+}$. We note that, as discussed in \cite{Bondarenko2017}, the emission of the 468.6\,nm  He$^+$ spectral line is from a transition from the n = 4 to n = 3 quantum state, and repopulation of the n = 4 state from the ground state requires collisional excitation by an electron with energy > 51 eV. This means that it is visible only when the He$^+$ ions interact with energetic electrons and therefore this emission line acts as a marker of collisionless energization of the background ions \cite{Bondarenko2017_2}. It clearly shows that the formation of the secondary diamagnetic cavity is spatially correlated with the emergence of a large blob of excited He$^+$ ions at the upper edge of the LPP, though the two structures remain separate and do not overlap. Measurements from the emissive probe do not show evidence of a persistent upward electrostatic field near the lower end of the $He^{+}$ blob that could account for the observed transverse ion displacement. This suggests that the field responsible for this transverse motion is most likely electromagnetic in nature. The C$^{4+}$ self emission images (lower row) show that while at earlier times, the LPP experiences a quasi-ballistic expansion, the inward magnetic pressure associated with the diamagnetic cavity limits its transverse expansion and leads to self-focusing and a mainly forward motion of the debris plasma. Additionally, the increased density compared to previous work\cite{Dorst2025} leads to stronger self-focusing of the LPP as well as stronger coupling with the background ions, with a much larger and persisting $He^{+}$ blob forming. 
 
In order to characterize the collisionless coupling leading to the formation and energization of the blob, the transverse (along y) velocity distribution of the excited He$^+$ ions was measured using Doppler spectroscopy on the 468.6\,nm line. As the fiber-probe is below the plasma looking up, light self-emitted by ions with a transverse velocity will be Doppler-shifted by $\Delta \lambda = V_y \lambda_0$/c , where $V_y$ is the velocity along the line of sight, $c$ is the speed of light, $\lambda_0$ is the rest wavelength. Therefore, positive $V_y$ will lead to a redshift and negative $V_y$ to a blueshift of the line self-emission. As shown in Fig. \ref{fig:fig4}(b), in early times, at $t = 600$ns and x=6\,cm, corresponding to the blob position, the line emission spectra are significantly redshifted, with a displacement of the peak of the spectrum of $\Delta \lambda = 0.07$~nm corresponding to a transverse velocity of $V_y=45\pm25$\,km/s, but a significant portion of the ions have an even higher velocity, as the half-width half-maximum of the spectra corresponds to $V_{+50\%} = 75$\,km/s. After $t=600$\,ns, the peak of the emission spectrum shift back towards lower wavelengths, indicating that the average velocity along y is decreasing. By $t=1400$~ns, the emission becomes  blueshifted relative to the reference spectrum, indicating that the ions are moving downward.

\begin{figure}[h]
\includegraphics[width=1\linewidth]{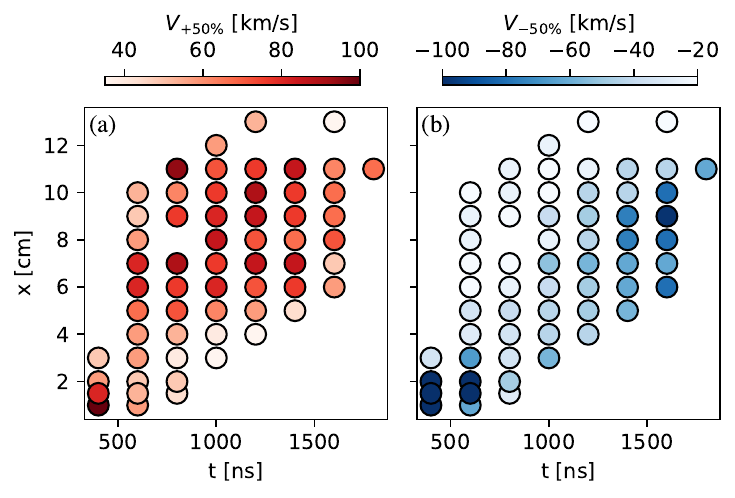}
\caption{\label{fig:fig5} (a) He$^+$ ions positive transverse velocity at half-maximum $V_{+50\%}$ and (b) Negative transverse velocity at half-maximum $V_{-50\%}$ obtained from Doppler spectroscopy.  }
\end{figure}

This observed behavior aligns with Larmor coupling, leading to the transverse acceleration of the He$^{+}$ blob, as illustrated in Fig.~\ref{fig:fig4}(b). The cross-field ion current $J_i$ generated by the laser-plasma expansion induces a laminar electric field $E_{\text{lam}}$ directed upward. This field accelerates background He$^{+}$ ions along the $y$-axis, initially causing the observed redshift. The ions then begin to gyrate in the background magnetic field, progressively converting their vertical velocity into a longitudinal component, as evidenced by the decrease in transverse velocity measured via Doppler spectroscopy. Specifically, the ions start acquiring a negative velocity at $t \sim 1400$~ns, corresponding to a quarter of the gyroperiod of He$^{+}$ ions in the background magnetic field ($\tau_i = 3.25 \,\mu$s), after being accelerated upward at $t = 600$~ns. 

\begin{figure*}[ht!]
\includegraphics[width=0.8\linewidth]{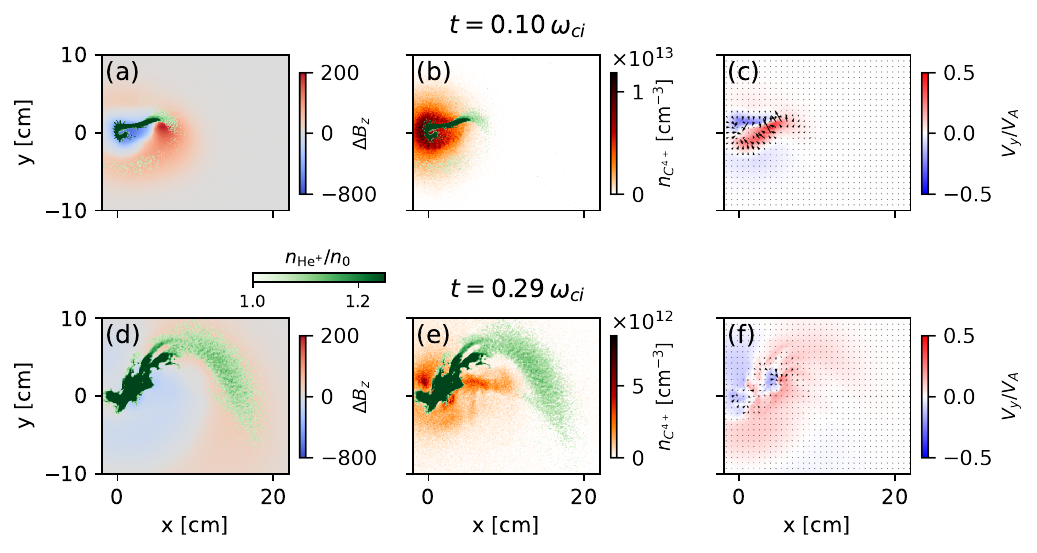}
\caption{\label{fig:fig6} \textit{PIC Simulations.} (Left) Variation of the z component of the magnetic field $\Delta B_z$ (cool-warm) and $He^{+}$ density normalized to background density showing only compressed density $n_{He^{+}}/n_0>1$ (green)  (Center) $C^{4+}$ ion density $n_{C^{4+}}$ from the LPP (orange) with previous $He^{+}$ density overlaid in green. (Right) Fluid velocity of the $He^{+}$ ions along y (red-blue) and arrows representing the strength and direction of the in-plane electric field, for simulation times $t = 0.10 \omega_{ci}$ (a-c) and $t = 0.29 \omega_{ci}$ (d-f).}
\end{figure*}

\begin{figure*}[ht!]
\includegraphics[width=0.9\linewidth]{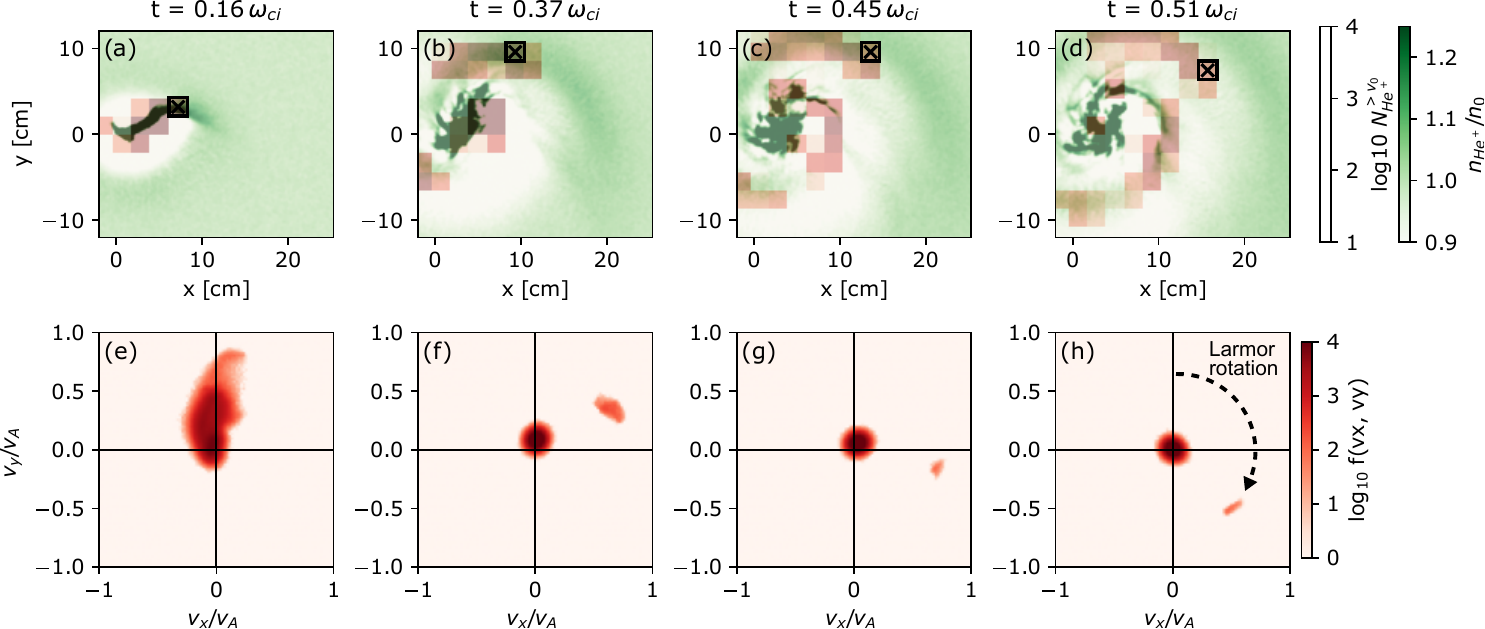}
\caption{\label{fig:fig7} \textit{Phase-space} (a)-(d) Evolution at different simulation times of the density of the background $He^+$ ions (Green) and log10 of the number of $He^+$ ions with velocity larger than $v_0^+ = 0.6V_A$  in each element of the spatial grid on which the phase-space is computed (Red).  (e)-(f) Phase-spaces $\log_{10} f(vx,vy)$, at the position of the phase-space spatial grid marked by the crossed box in the top panels.   }
\end{figure*}

Additionally, as the emission of the ions becomes blueshifted at later times, a significant portion of the spectrum remains redshifted, resulting in an overall broadened spectrum. This suggests that while the first ions begin to rotate, new ions continue to be accelerated vertically by the laminar electric field. To account for this effect, we determine from each spectrum the positive and negative velocities, $V_{+50\%}$ and $V_{-50\%}$, corresponding to the half-width at half-maximum (HWHM) on each side. These values provide insight into the extent of ion acceleration in both directions. Taking advantage of the high-repetition rate, we were able to acquire Doppler spectra for different positions $x \in \{ 1.5, 13 \}~\text{cm}$ and times $t \in \{ 400, 1800 \}~\text{ns}$, and half-maximum positive and negative velocities for each point are represented in Fig\,\ref{fig:fig5}. Initially, we observe an important spectral broadening corresponding to a high velocity of the He$^{+}$ ions in both directions at early times ($\leq600$\,ns) and close to the target ($x\leq2$\,cm). This localized energization of the background plasma is not associated with Larmor coupling, but is most likely due to strong fields generated at the target surface during the laser interaction, and will not be discussed further in this paper.  At positions further away from the target, the trend indicated in Fig. \ref{fig:fig4} is confirmed. Initially, the background ions acquire a significant upward velocity—with $V_{+50\%}$ reaching around 100\,km/s—and no downward velocity is observed. Later, as the He$^+$ ions gyrate, large negative velocities of similar magnitude are observed for $t>1400$~ns, and at progressively larger $x$ positions as the blob propagates forward. This later increase in downward momentum, along with a moderate decrease in $V_{+50\%}$, provides clear evidence of Larmor rotation of the ions.

\section{PIC simulations and discussion}
\label{sec:PIC}
In order to gain more insight on the formation of the $He^{+}$ blob and its subsequent collisionless coupling to the LPP through Larmor coupling, we carried out fully kinetic PIC simulations using the code \textit{VPIC} \cite{Bowers08,Le2021}. The simulations are realized in a 2D3V geometry in the (XY)-plane at $z=0$, in a square domain of size $L_x = L_y = 6\,d_i$, discretized on a uniform grid with $N_x = N_y = 600$ cells with periodic boundary conditions, with $n_{ppc} = 1000$ macroparticles per cell in the uniform background plasma. The ion-to-electron mass ratio is set to $m_{He}/m_e = 50$ and the ratio of the plasma frequency and the electron gyrofrequency is set to $\omega_{pe}/\omega_{ce}=2$ to reduce computational costs. Key parameters that are matched to the experimental data include the Alfvenic Mach number of the expanding carbon plasma $M_A = 2.1$ and total energy in the debris ions, which sets the maximal size of the magnetic cavity. The expanding 
LPP carbon ions are initialized with density and velocity profiles collimated in the forward direction similar to previous simulation studies \cite{Bondarenko2017}, and they include a faster ($M_A=2.1$) population of $C^{4+}$ ions  with $\sim 2/3$ of the initial kinetic energy and a slower ($M_A=0.7$) population of  $C^{2+}$ ions with the remaining energy. 

Figure \ref{fig:fig6} shows the evolution of different quantities associated with the LPP expansion in background plasma in the simulation, analogous to quantities measured in the experiment. Figures \ref{fig:fig6}(a,d) show that the LPP expansion generates a diamagnetic cavity, while leading to a compression of the magnetic field at the leading edge as was observed experimentally. The strong magnetic field gradients of the diamagnetic cavity lead to the collimation of the LPP and the formation of a jet-like structure at its leading edge (see Fig. \ref{fig:fig6}b,e). Two different background $He^+$ structures are formed. First, the strong inward electric fields associated with the magnetic field gradients around the edge of the diamagnetic cavity (see Fig. \ref{fig:fig6}c) push and concentrate background ions toward the center, leading to a dense but slow population of $He^+$, which is in good accordance with the strong $He^+$ emission observed in the experiment at early times, near the center of the cavity (see Fig. \ref{fig:fig3}e-f). Another population of background ions is accelerated upward by the vertical $\mathbf{J_i} \times \mathbf{B}$ electric field resulting from the cross-field forward motion of the collimated LPP at its leading edge (see Fig. \ref{fig:fig6}c). This second population is responsible for the blob as discussed in the experimental section. 

To characterize the coupling of the LPP to the background plasma, we examine the phase-space structure of the background $He^{+}$ particles in the PIC simulation. To obtain a sufficient statistical sample, we bin particles on a coarser $20 \times 20$ spatial grid, and we construct a reduced phase-space histogram in the $v_x$---$v_y$ plane (perpendicular to the background magnetic field) using a $100 \times 100$ velocity grid spanning from $-v_A$ to $v_A$ in each direction. We also define the number of $He^+$ ions with a velocity over $v_0^+ = 0.6V_A$ as $N^{>v_0^+}_{He^+}$ to quantify the energization of the background ions through collisionless coupling with the LPP [see Figs.~\ref{fig:fig7}(a-d)]. This phase-space data is summarized in Fig. \ref{fig:fig7} and shows that a population of $He^+$ ions near the upper edge of the cavity is energized and initially pushed upward. As shown in Fig. \ref{fig:fig7}e the vertical electric field at the tip of the LPP accelerates a large population of background $He^{+}$ ions in  the $+y$ direction, and numerous background ions have $v_y>0.6v_A$ in Fig.~\ref{fig:fig7}a. As time advances, the background magnetic field $\mathbf{B_0} = B_0 \mathbf{u_z}$ rotates this energized population in phase-space, and converts this upward $+y$ momentum into forward $+x$ momentum. As the Larmor rotation continues, some of the momentum is transferred back in the downward $-y$ direction. This is in good accordance with the experimental observation of an initial redshift and a later blueshift of the line emission of $He^+$, and it is a clear signature of the Larmor coupling mechanism.  \par
While the simulations reproduce quite well the general dynamics of the ions and their collisionless coupling, we note that we do not observe as clearly a secondary diamagnetic cavity associated with the accelerated blob detaching from the main one in the simulations. At $t=0.1\omega_{ci}$, a small protrusion is starting to form associated with the $He^{+}$ blob, but it does not fully develop later on, as the density of the coupled blob remains relatively low. This discrepancy may arise from the simplified velocity distribution used to initialize the debris plasma, which is necessarily less realistic than the experimental LPP, and from the simulation’s 2D geometry.

\section{Conclusions}

We reported laboratory measurements of collisionless Larmor coupling during the expansion of a laser-produced plasma (LPP) into a strongly magnetized, low-$\beta$ ambient plasma on LAPD. Using a high-repetition-rate laser, we performed systematic spatial and temporal scans of the interaction with complementary diagnostics: time-resolved magnetic field and plasma potential mapping in the $XY$ plane, narrow-band self-emission imaging of both background He$^{+}$ and debris C$^{4+}$, and Doppler spectroscopy of the He$^{+}$ 468.6\,nm line to directly access the transverse ($y$) velocity distribution of energized background ions.

The measurements show that the LPP forms a diamagnetic cavity whose strong magnetic pressure gradients collimate the debris flow. In the presence of a background plasma, we observe the appearance of a distinct, excited He$^{+}$ blob at the upper edge of the LPP, spatially correlated with the development of secondary magnetic-field structure. Doppler spectroscopy provides direct evidence of collisionless ion energization consistent with Larmor coupling: at the blob location, the He$^{+}$ emission line is initially redshifted, indicating upward acceleration, and later evolves toward blueshifted emission at delays comparable to a fraction of the He$^{+}$ gyroperiod. The coexistence of red- and blueshifted components at late times implies a sustained source of upwardly accelerated ions while earlier ions undergo gyromotion, producing a broadened line. To interpret these observations, we performed VPIC simulations initialized to match the experimental magnetization and super-Alfv\'enic conditions. The simulations reproduce the formation of a diamagnetic cavity, the collimation of the debris ions, and the emergence of an energized population of background He$^{+}$ ions near the leading edge of the LPP, that subsequently rotate in phase-space through Larmor coupling. 

These results provide a detailed experimental characterization of blob formation and ion energization driven by collisionless Larmor coupling in a super-Alfv\'enic, magnetically dominated regime, complementing previous work \cite{Bondarenko2017,Dorst2025} with extensive characterization of the Larmor coupling mechanism.

\begin{acknowledgments}
This work was supported by the Defense Threat Reduction Agency and Lawrence Livermore National Security LLC under Contracts No. B649519, B661613, and B671327, and by the National Science Foundation under Contract No. NSF PHYS-2320946. The experiments were performed at the UCLA Basic Plasma Science Facility (BaPSF), which is a collaborative research facility supported by the U.S. Department of Energy, Office of Science, Fusion Energy Sciences program, and the National Science Foundation.
\end{acknowledgments}

\section*{Data Availability Statement}

The data that support the findings of this study are available from the corresponding author upon reasonable request.

\bibliographystyle{apsrev4-2}
\bibliography{bib}

\end{document}